\documentclass[prl,twocolumn,superscriptaddress,showpacs]{revtex4}
\usepackage{times}
\usepackage{amsmath}
\usepackage{amsthm}
\usepackage{amssymb}
\usepackage{amsbsy}
\usepackage{graphicx}
\usepackage{pstricks}

\begin{document}
\bibliographystyle{apsrev}
\title{Charge degrees in the quarter-filled checkerboard lattice}

\author{Frank Pollmann}
\affiliation{Max-Planck-Institute for the Physics of Complex Systems, D-01187 Dresden, Germany}

\author{Joseph J. Betouras}
\affiliation{Instituut-Lorentz for Theoretical Physics, P.O. Box 9506,
NL-2300RA Leiden, The Netherlands}
\affiliation{School of Physics and Astronomy, 
Scottish Universities Physics Alliance, 
University of St Andrews, North Haugh KY16 9SS, UK}
\author{E. Runge}
\affiliation{Technische Universit\"at Ilmenau, Institut f\"ur Physik, 98684 Ilmenau, Germany}
\author{Peter Fulde}
\affiliation{Max-Planck-Institute for the Physics of Complex Systems, D-01187 Dresden, Germany}

\begin{abstract}
For a systematic study of charge degrees of freedom in lattices with geometric frustration, we consider spinless fermions on the checkerboard lattice with nearest-neighbor hopping $t$ and nearest-neighbor repulsion $V$  at quarter-filling. An effective Hamiltonian for the limit $|t|\ll V$ is given to lowest non-vanishing order by the ring exchange ($\sim t^{3}/V^{2}$). We show that the system can equivalently be described by hard-core bosons and map the model to a confining $U(1)$ lattice gauge theory.
\end{abstract}

\maketitle

\date{\today}

\newcommand{\orients}{\begin{picture}(6,6)(1,1) \includegraphics[height=3mm, width=3mm]{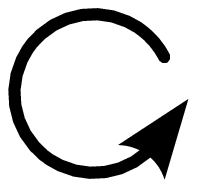} \end{picture}}

\newcommand{\smallrech} {\begin{picture}(6,4)(0,0) \includegraphics[height=1mm, width=2mm]{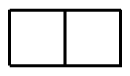} \end{picture}}
\newcommand{\smallrecv} {\begin{picture}(4,6)(0,0) \includegraphics[height=2mm, width=1mm]{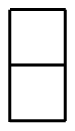} \end{picture}}

\newcommand{\emptya}{\begin{picture}(20,10)(0,5) \includegraphics[height=5mm, width=7mm]{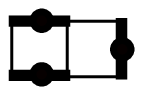} \end{picture}}
\newcommand{\emptyb}{\begin{picture}(20,10)(0,5) \includegraphics[height=5mm, width=7mm]{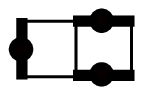} \end{picture}}

Recently, there has been enormous interest in the properties of quantum magnets on lattices with frustrated geometries where the number of classical ground state configurations increases exponentially with the number of sites. The arguably most spectacular phenomenon in this field is fractionalization of quantum numbers (e.g., spinons), see Refs. \cite{balents2002,diep2005} and citations therein. Now, attention  shifts to charge degrees of freedom. A model on the checkerboard lattice has been introduced in which an added fermion can decay into two mobile pieces which carry the fractional charge of $e/2$ each \cite{fulde2002}. Numerical studies of this model have shown that the fractional charges are weakly confined and form bound pairs of large spatial extent. These are expected to lead to interesting physical effects \cite{runge2004,pollmann2006}. In the present work, a related model, namely the checkerboard lattice on which one-quarter of the sites are occupied by strongly interacting spinless fermions, is considered. We dicuss an effective Hamiltonian that describes the low-energy physics. This Hamiltonian has a $U(1)$ gauge invariance which we use to relate the problem at hand to the compact quantum electrodynamics in $2+1$ dimensions \cite{polyakov1977}. 

\begin{figure}
\begin{center}\begin{tabular}{cc}
(a)\includegraphics[height=28mm]{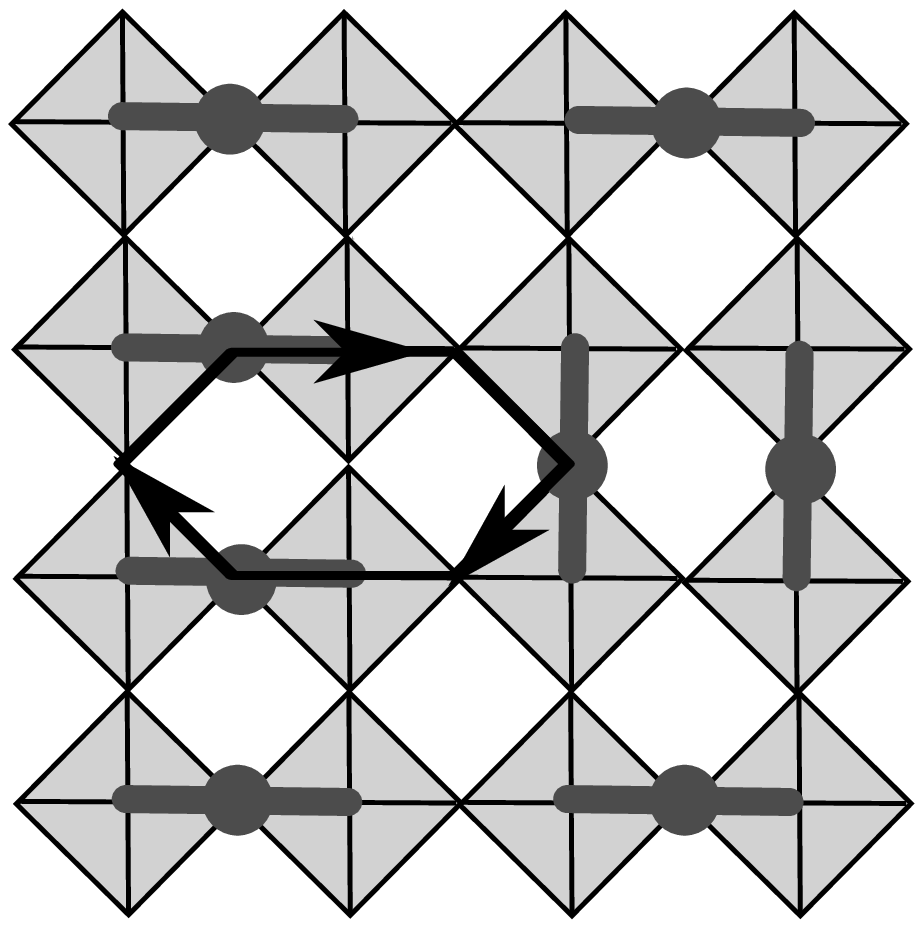}&
~~(b)\includegraphics[height=28mm]{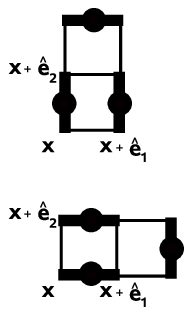}
\end{tabular}\end{center}
\caption{(a) Square lattice connecting the centers of the crisscrossed squares of the checkerboard lattice with links. One exchange process around a hexagon is shown explicitely. (b) Labeling of links on flipable double-plaquettes at position $\mathbf x$ with unit vectors $\hat{e}_1$ and $\hat{e}_2$.\label{labeling}}
\end{figure}

Our starting point is the Hamiltonian of spinless fermions in the checkerboard lattice with nearest-neighbor hopping amplitude $t$ and a strong repulsive interaction $V$ on neighboring sites. The low-energy manifold at quarter-filling in the limit of strong interactions ($|t|\ll V$) is given by those configurations which have exactly one fermions on each crisscrossed square. This local constaint corresponds to the so-called tetrahedron rule, which applies in the half-filled case \cite{fulde2002,anderson1956}. The low-energy manifold of states fullfilling the local constraint can alternatively be represented by a model on the square lattice connecting the centers of the crisscrossed squares. The particles sit in the middle of the dimers (see Fig.~\ref{labeling}~(a)). The condition of one fermions per crisscrossed square is being translated into the requirement of one dimer touching each site. Hence, states satisfying this constraint are represented by hard-core dimers on the square lattice. Notice that this model is very reminiscent of  quantum dimer models on the square lattice \cite{rokshar1988}.\\
In lowest non-vanishing order in $t/V$, the Hamiltonian for spinless fermions on a quarter-filled checkerboard lattice can be reduced to the following effective Hamiltonian \cite{runge2004}:
\begin{eqnarray}
H_{\mbox{eff}}=-&g&\sum_{\{\smallrech,\smallrecv\}}\left( \Big|\emptya\Big\rangle\Big\langle\emptyb\Big|+\Big|\emptyb\Big\rangle\Big\langle\emptya\Big|\right ). \label{eq:ring-exchange1}\label{Heff}
\end{eqnarray}
The sums are performed over all polygons of perimeter six, and $g\sim t^3/V^2$. The dots indicate particle locations. There is a difference with respect to the 'bosonic' quantum dimer model \cite{rokshar1988} resulting from the fermionic statistics: The smallest resonating plaquette has perimeter six, rather than four. It is noticeable that \textit{no Fermi sign} problem remains. Only those processes that have the same overall sign will occur in the ground state of the quarter-filled case \cite{runge2004}. In passing we mention that the overall sign of the ring-exchange amplitude $g$ in (\ref{Heff}) can be chosen arbitrarily.\footnote{The effective Hamiltonian (\ref{Heff}) changes the number $N_{SL}$ of fermions on a certain sublattice always by two, thus a multiplication of all configurations with the factor $i^{N_{SL}}$  changes the sign of the ring-exchange amplitude $g$.}

The effective Hamiltonian (\ref{Heff}) conserves the number of fermions on each crisscrossed square. Consequently, the number of dimers touching each site of the lattice connecting the centers of the crisscrossed squares is not changed. This conservation generates a $U(1)$ gauge invariance, as it is usually the case in dimer models  \cite{fradkin1990}. The gauge structure suggests that we can gain further insight with respect to low-energy excitations by writing our model as an $U(1)$ lattice gauge theory. The usefulness of this approach has already been shown for the quantum dimer model (QDM) on the square lattice \cite{fradkin1990} as well as for three dimensional spin systems \cite{hermele2004}.

Define for each link $(\mathbf{x},\mathbf{x}+\hat{e}_{j})$ of the square lattice an integer variable $n_{j}(\mathbf{x})$, where $\mathbf{x}$ denotes
the coordinates of a lattice site and $\hat{e}_{j=1,2}$ are unit vectors along the axes as shown in Fig.~\ref{labeling}(b).  The states $|\{n_{j}(\mathbf{x})\}\rangle$ span an enlarged Hilbert space which has integer numbers for the links instead only zero or one. We can consider the states $|\{ n_{j}(\mathbf{x})\}\rangle$ as eigenstates of quantum rotor operators $\hat{n}_{j}(\mathbf{x})$ with eigenvalues $n_{j}(\mathbf{x})$.

In order to express the effective Hamiltonian in terms of the $\hat{n}_j(\mathbf x)$, we introduce phases $\hat{\phi}_j(\mathbf{x})\in [0,\pi)$  on the links which are canonical conjugate to $\hat{n}_{j}(\mathbf{x})$. Using the fact that $\exp\left[ \pm i \hat{\phi}_j(\mathbf{x})\right ]$ act as ladder operators, we can write 
\begin{eqnarray}
\mathcal{H}_{\mbox{eff}}= U {\sum_{\mathbf{x}, j}}\left (  \hat {n}_j(\mathbf x)-\frac{1}{2}\right) ^2 -2g\sum_{\{\smallrech,\smallrecv\}}\cos \left[\sum \pm \hat \phi \right]. \label{Hgauge}
\end{eqnarray}
Here, the argument of the cosine term contains the sum over phases $\hat {\phi}_j(\mathbf{x})$ with alternating signs around the polygons of perimeter six (double plaquettes). In the limit $U/g\rightarrow \infty$ Eq.~(\ref{Hgauge}) is a faithful representation of the effective Hamiltonian (\ref{Heff}). 

We introduce staggered gauge and electric fields on the bipartite square lattice $\mathbf{x}=(x_{1},x_{2})$ by
\begin{eqnarray*}
\hat{A}_{j}(\mathbf{x}) & = & (-1)^{x_{1}+x_{2}}\hat{\phi}_{j}(\mathbf{x})\\
\hat{E}_{j}(\mathbf{x})& = &(-1)^{x_{1}+x_{2}}\left(\hat{n}_{j}(\mathbf{x})-\frac12\right),\\
\end{eqnarray*}
The local constraint that each site is touched by exactly one dimer reads
\begin{eqnarray}
\left( \Delta_{j}\hat{E}_{j} \left( \mathbf{x} \right) - \rho(\mathbf{x}) \right) |\mbox{Phys.}\rangle = 0,\ \rho(\mathbf{x})=(-1)^{(x_1+x_2+1)},
\label{gauss}
\end{eqnarray}
where the lattice divergence is defined as
\begin{eqnarray*}
\Delta_{j}\hat{E}_{j}\left(\mathbf{x}\right) \equiv \hat{E}_{1}(\mathbf{x})-\hat{E}_{1}(\mathbf{x}-\mathbf{e}_{1})+\hat{E}_{2}(\mathbf{x})-\hat{E}_{2}(\mathbf{x}-\mathbf{e}_{2}). \end{eqnarray*}
The constraint is now reflected by the standard Gauss' law (\ref{gauss}) in the presence of a staggered background charge density. The Hamiltonian (\ref{Hgauge}) in staggered variable reads
\begin{eqnarray}
\mathcal{H}_{\mbox{eff}}=U{\sum_{\mathbf{x}, j}}\hat{E}_{j}^2(\mathbf{x})-2g\sum_{\mathbf{x}} \cos \left[\sum_{\orients} \hat{A}_{j}\left(\mathbf{x}\right) \right].\label{Hqed}
\end{eqnarray}
Here, the argument of the cosine term denotes the oriented sum of staggered vector potentials $\hat{A}_{j}\left(\mathbf{x}\right)$  around double plaquettes. The vector $\mathbf{x}$ labels the site in the left bottom as shown in Fig.~{\ref{labeling}}. Eq.  (\ref{Hqed}) has similarities with the Hamiltonian of the compact quantum electrodynamic (QED) in $2+1$ dimensions in which the considered charges correspond to fractional charges of $e/2$ \cite{polyakov1977}. Polyakov showed for the compact QED in $2+1$ dimension that it has an unique and gapped ground state. Two charges are confined and the energy grows linearly with the distance between the two charges. Our model shows important differences. The fields $E_{j}(\mathbf{x})$ are half integers instead of integers and the constraint selects configurations with a background charge $\rho(\mathbf{x})$. This leads to a frustration which is reflected by the macroscopic degeneracy of the classical ground states. Furthermore, the definition the flux in the cosine term differs. 

Formulation (\ref{Hqed}) provides an excellent starting point for further systematic investigations. The plaquette duality transformation allows to map the Hamiltonian to a height model and to use path integrals for a detailed study of the ground-state as well as low-energy excitations \cite{fradkin1990}.

\end{document}